\documentclass[10pt,english, aps,superscriptaddress,showpacs,floatfix,prl,lengthcheck,floatfix,nofootinbib,twocolumn]{revtex4-2}

\usepackage[T1]{fontenc}
\usepackage[utf8]{inputenc}
\usepackage{amsmath}
\usepackage{amssymb}
\usepackage{graphicx}
\PassOptionsToPackage{normalem}{ulem}
\usepackage{ulem}
\usepackage{graphicx} 
\usepackage{float}
\usepackage[utf8]{inputenc}
\makeatletter

\usepackage[T1]{fontenc}
\usepackage[utf8]{inputenc} 

\usepackage{braket}

\usepackage{amsmath}

\usepackage{color}
\usepackage{babel}
\usepackage{verbatim}
\usepackage{amsmath}
\usepackage{amssymb}
\usepackage{graphicx}
\usepackage[pdfusetitle,colorlinks,linkcolor=blue,citecolor=blue,urlcolor=blue]{hyperref}
\usepackage{microtype}

\makeatother

\usepackage{babel}
\begin{document}
\title{Effects of Zero-Point Motion in the High Harmonic Generation Spectrum of Solids}
\author{Aday C\'ardenas}
\affiliation{\emph{Instituto de Ciencia de Materiales de Madrid (ICMM), Consejo
Superior de Investigaciones Científicas (CSIC), Sor Juana Inés de
la Cruz 3, 28049 Madrid, Spain}}
\author{David N. Purschke}
\affiliation{\emph{Laboratory for Laser Energetics, University of Rochester, Rochester, New York 14623, United States}}
\author{Graham G. Brown}
\affiliation{\emph{Max Born Institute, Max-Born-Straße 2A, 12489, Berlin, Germany}}
\author{Pablo San-Jose}
\affiliation{\emph{Instituto de Ciencia de Materiales de Madrid (ICMM), Consejo
Superior de Investigaciones Científicas (CSIC), Sor Juana Inés de
la Cruz 3, 28049 Madrid, Spain}}
\author{Rui E. F. Silva}
\affiliation{\emph{Instituto de Ciencia de Materiales de Madrid (ICMM), Consejo
Superior de Investigaciones Científicas (CSIC), Sor Juana Inés de
la Cruz 3, 28049 Madrid, Spain}}
\affiliation{\emph{Max Born Institute, Max-Born-Straße 2A, 12489, Berlin, Germany}}
\author{\'Alvaro Jim\'enez-Gal\'an}
\email{alvaro.jimenez@csic.es}
\affiliation{\emph{Instituto de Ciencia de Materiales de Madrid (ICMM), Consejo
Superior de Investigaciones Científicas (CSIC), Sor Juana Inés de
la Cruz 3, 28049 Madrid, Spain}}
\affiliation{\emph{Max Born Institute, Max-Born-Straße 2A, 12489, Berlin, Germany}}

\begin{abstract}
The interpretation of high-harmonic generation (HHG) in solids typically relies on phenomenological dephasing times far shorter than what is expected from microscopic scattering processes. Here we show that zero-point fluctuations associated with optical phonons naturally suppress long-range electronic coherences and generate clean harmonic spectra without introducing ad-hoc decoherence parameters. Using a 1D semiconductor composed of two distinct sites per unit cell and realistic phonon amplitudes, we demonstrate that random per-site optical-phonon jitter reproduces the spectral sharpening typically attributed to ultrafast $T_2$ dephasing. In contrast, acoustic phonons and local strain, whose distortions are correlated over nanometer scales, produce negligible spectral cleaning. We further show that such long-range site coherence leads to carrier-envelope-phase-dependent effects in the HHG spectrum driven by long pulses, but these effects collapse once optical-phonon-induced decoherence is included. Our results (i) identify optical zero-point motion as a key mechanism governing coherence in solid-state HHG, (ii) demonstrate that it can be qualitatively modeled in periodic solids through site-distance-dependent dephasing, and (iii) suggest that CEP-resolved measurements can probe electronic coherence lengths and atomic fluctuations in crystalline materials.
\end{abstract}
\maketitle

\section{Introduction}
The first observation of high harmonic generation (HHG) from a bulk crystal opened the way to a plethora of investigations on condensed matter systems on the femtosecond to attosecond timescale~\cite{Ghimire2011, Vampa2015,Luu2015,Higuchi2017,Schoetz2019,Lakhotia2020,Uzan-Narovlansky2022,Heide2024,Heide2024a}. These timescales are particularly relevant since they correspond to those of coherent electron dynamics inside atoms, molecules and condensed matter~\cite{Krausz2009,Kruchinin2018}. The strong laser field used to generate the high-order harmonics can guide electron motion in the solid coherently~\cite{Vampa2015nat}, opening the way to the manipulations of its quantum properties and to the study of electronic decoherence process in the solid~\cite{Jr2019,Nie2023,Mitra2024,Tyulnev2024,Freudenstein2022,Zong2023,Uzan-Narovlansky2024}. Yet, ever since the first theoretical description of the HHG process in crystals, the agreement between theory and experiment has relied on the use of phenomenological decoherence parameters with values that are not physically justified. Vampa et al. showed that to fit experiments it was necessary to introduce an extremely short dephasing time ($T_2\approx4~$fs) into the semiconductor Bloch equations in order to reduce the number of interfering trajectories which will otherwise lead to a noisy spectrum with no clear harmonics~\cite{Vampa2014}. The physical origin of such short dephasing times is an active area of research~\cite{Yue2022}. 

Floss et al. argued that incorporating the effects of macroscopic light propagation through the solid allows to obtain a cleaner harmonic spectrum with a more modest dephasing time ($T_2=10~$fs)~\cite{Floss2018}. While such effect most likely plays an important role, the observation of clean high harmonics from 2D materials and in reflection geometries~\cite{Liu2017}, where the propagation effects should be unimportant, shows that it is not the full picture. Du and Ma later showed that acoustic lattice vibrations in a monoatomic 1D carbon chain, implemented by independent random fluctuations of each bond, could be mapped to an effective $T_2$ dephasing~\cite{Du2022}. For a Debye temperature of $\Theta=150$~K they obtained an effective $T_2\sim 25$~fs at room temperature, which is still higher than that required to obtain a clean HHG spectrum. Additionally, acoustic vibrations are generally highly correlated through distances of several nanometers, so that an uncorrelated per-bond jitter will likely overestimate their effect. Similarly, through time-dependent density functional theory calculations in a supercell, Freeman et al. showed that atomic motion due to thermal effects suppresses very high harmonics and increases their peak-to-peak depth, and suggested cooling as a way to increase the overall harmonic yield~\cite{Freeman2022}. Experiments have however still not shown a clear temperature dependence of harmonics besides those occurring at phase transitions~\cite{Alca2022}. It is worth mentioning that nuclear motion has been well-studied in the context of HHG from molecular gases, with works including full quantum treatment of the nuclear wavepacket~\cite{Patchkovskii2009,Silva2013,Patchkovskii2017,Catoire2014, Nisoli2017,Richter2019}, showing their effect on light molecules~\cite{Baker2008,Smirnova2009,Lara-Astiaso2016}. In the context of solid-state HHG, several works have shown that HHG can be exploited to probe coherently driven phonon dynamics. Luu et al. demonstrated that time-resolved HHG spectroscopy can resolve excited lattice motion and even multi-phonon scattering pathways in solids~\cite{Zhang2024}, while Neufeld et al. showed that phonons driven by a terahertz pump pulse modulate and partially wash out the high-harmonic emission from a delayed infrared probe~\cite{Neufeld2022}. These studies reveal the high sensitivity of HHG to lattice motion when phonons are externally excited. 

Recently, Brown et al. introduced a dephasing rate with a quadratic dependence on site distance to mimic macroscopic effects~\cite{Brown2024}. This was initially justified on the grounds that focal-averaging removes trajectories between sites which accumulate more than a $2\pi$ phase, although recent work by Purschke et al. has shown that such parametrization can reproduce also the effects of medium-range disorder on the HHG spectrum of an amorphous crystal~\cite{Purschke2025}, which is similar to limiting the excursion length that a recombining electron trajectory can take~\cite{Mondal2023}. A complementary approach by Orlando et al. demonstrated that static disorder, introduced through an Anderson-type impurity potential, generates an effective dephasing that damps long interband trajectories and cleans the harmonic spectrum if sufficiently strong disorder is present~\cite{Orlando2020}. Since this mechanism relies on extrinsic impurity scattering, it cannot account for clean HHG spectra observed in high-quality crystals and 2D materials, where disorder is minimal. Both in the case of a $T_2$ and a distance-dependent dephasing, an ad-hoc parameter is used to fit the experiment, and removing long-range coherences without a microscopic justification may prove problematic if one wishes to describe excitonic wavefunctions with a large radius~\cite{Jensen2024, Molinero2024}. 

Here, we show that zero-temperature quantum jitter of atoms due to optical phonons, occurring naturally in all HHG experiments, can lead to a clean harmonic spectrum of a crystal without the need to include dephasing terms ad-hoc, and that these effects are weakly dependent on temperature. To illustrate this, we consider a 1D semiconductor, modeled by a finite 1D chain composed of two distinct atomic sites (A and B) per unit cell and a total of 1199 atoms, which we show gives the same results as a fully periodic chain, and use realistic optical phonon frequency parameters to model the random atomic position fluctuations. Since large finite-system calculations are currently unfeasible for 2D and 3D systems, we compare the HHG spectrum obtained as a function of the average displacement with the HHG obtained from a periodic system as a function of a dephasing time $T_2$ and a site-distance-dependent dephasing as introduced by Brown et al.~\cite{Brown2024}. We conclude that the latter reproduces more faithfully the effect of optical phonon jitter. Using a correlated jitter model, we show that, in contrast to optical phonons, acoustic phonons and local strain will not generally produce a clean harmonic spectrum due to the long-range correlations of the bond distortions as compared to the typical lengths of electronic excursion paths ($\sim $~\AA\,- few nm). Finally, we discuss how the experimental carrier-envelope-phase (CEP) instability can lead to CEP-effects even for long pulses when the solid system has long-range coherences, opening a way to reconstruct such coherence range from CEP-stable experiments.

\section{Results\protect\label{sec:Results}}

\begin{figure}
    \centering
    \includegraphics[width=\linewidth]{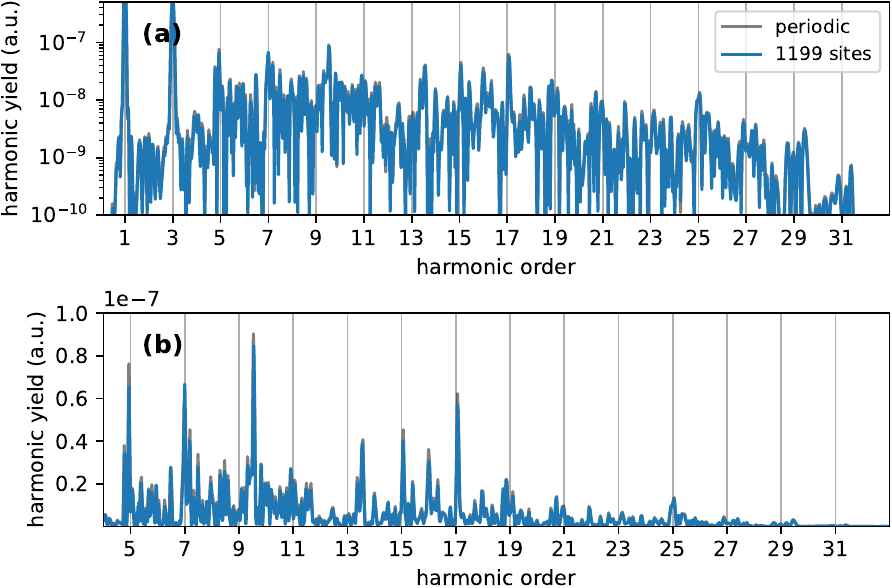}
    \caption{Comparison of the high-harmonic generation spectra between a periodic 1D chain of atoms and a finite chain of $N=1199$ atoms (see text for tight binding parameters). In both cases, atoms have fixed, well-defined positions. (a) Logarithmic scale. (b) Linear scale.}
    \label{fig:hhg_periodic}
\end{figure}

We first consider the pristine case in which the atoms are equally spaced, that is, $x_B-x_A = a_0/2$ and $q_n=0$. We simulate the interaction with a Gaussian electric field pulse of amplitude $E_0=0.005$~a.u. ($I=1$~TW/cm$^2$), frequency $\omega=0.387$~eV ($\lambda=3200$~nm) and full-width-at-half-maximum pulse duration of 80~fs, in the dipole approximation and in the length gauge. For the periodic system, we use the code~\cite{Molinero2025}, which exploits periodic boundary conditions. For the finite chain, we propagate the full dense density matrix of dimension $N \times N$, where $N$ is the number of atoms in the chain, with open boundary conditions using a Runge-Kutta 4 propagator. Fig.~\ref{fig:hhg_periodic}a,b shows the high-harmonic spectrum for the periodic and pristine finite-size chain with $N=1199$ atoms without including any dephasing term. Both give essentially the same result, meaning that the finite chain for $N=1199$ can be considered periodic. We note that we choose an odd total number of atoms to preserve the inversion symmetry. Due to the absence of dephasing, the HHG spectrum is noisy, with maxima appearing at positions other than those of the fundamental harmonics, in sharp contradiction to typical experimental solid-HHG spectra. The reason is similar to that of hyper-Raman forbidden lines predicted and observed in HHG from gas atoms~\cite{Millack1993,Moiseyev2003,Jimnez-Galn2017,Bloch2019}, and also those appearing in extended correlated systems~\cite{Lange2024}, which are further amplified in solid-state simulations due to the smaller energy scales.

\begin{figure}
    \centering
    \includegraphics[width=\linewidth]{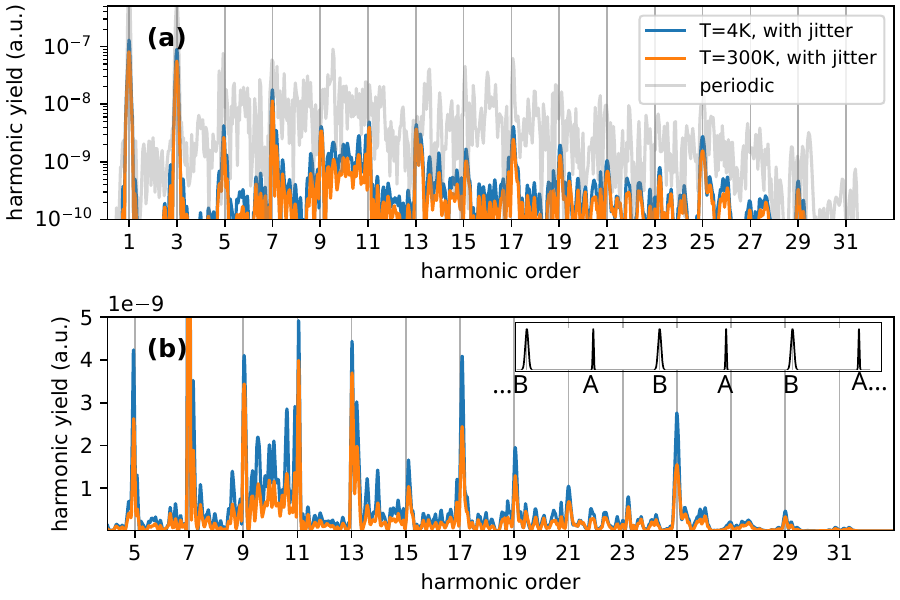}
    \caption{Comparison of the HHG spectra between the periodic chain (gray) and two finite chains of $N=1199$ atoms with optical-phonon-like site jitter at temperatures of $T=4$~K (blue) and $T=300$~K (orange). Coherent averaging was performed over 20 random simulations, which was enough to attain convergence due to self-averaging effects of the long chain. The inset in (b) illustrates the chain with the width of the Gaussian function representing the per-atom root-mean-squared displacements, alternating between the heavier (A) and lighter (B) atoms, and the center of the Gaussians corresponding to the equilibrium static positions. (a) Logarithmic scale. (b) Linear scale.}
    \label{fig:hhg_finite}
\end{figure}

\begin{figure*}
    \centering
        \includegraphics[width=0.9\textwidth]{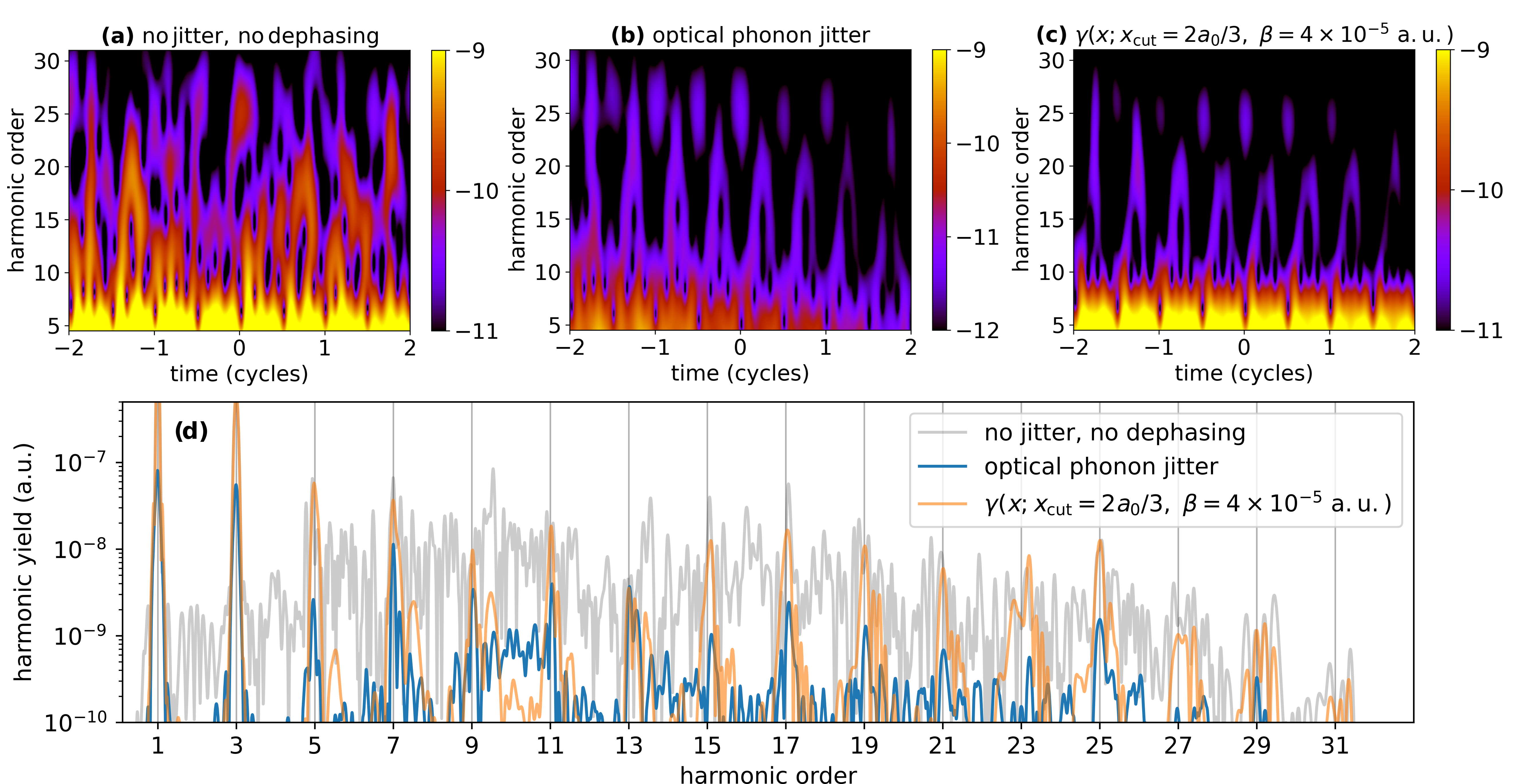}
    \caption{(a-c) Time-frequency plots of the current for a 1D chain of $N=1199$ atoms. a) Chain with no atomic jitter and no phenomenological dephasing terms, displaying multiple trajectories per cycle. (b) Chain with atomic jitter consistent with optical phonon amplitudes. (c,d) Chain with no atomic jitter and a distance-dependent dephasing term of  $\gamma(x; x_{cut}=2a_0/3,\beta=4\times 10^{-5}~\text{a.u.})$. (d) HHG spectrum for the case of panel a (gray curve), b (blue curve) and c (orange curve), in logarithmic scale.}
    \label{fig:gabor}
\end{figure*}

Using the finite chain with $N=1199$ atoms, we can now consider the effect of random per-site jitter calibrated to optical phonon displacements. Details can be found in Appendix~\ref{sec:methods}. In particular, we choose parameters similar to those of monolayer MoS$_2$: $a_0 = 3.17$~\AA, $\varepsilon_B-\varepsilon_A = 1.9$~eV, $\omega_{\text{phon}}=49$~meV, $m_{A}=1.59\times10^{-25}$~kg, $m_{B}=5.32\times10^{-26}$~kg, $t_0=2.7$~eV, and $\gamma=3.37$ (these last two were taken similar to graphene due to the more abundant literature)~\cite{Lee2010,Pereira2009}. The top inset in Fig.~\ref{fig:hhg_finite}b depicts the root-mean-square (RMS) deviation of the atomic displacement calculated with these parameters relative to the equilibrium atomic distances. The HHG spectrum, shown in Fig.~\ref{fig:hhg_finite}a,b, now displays clear peaks at the harmonic energies.  We stress that no ad-hoc dephasing mechanism was introduced here either. Temperature plays only a minor role, as can be seen by comparing the HHG spectra for $T=4$~K and $T=300$~K; most of the atomic displacement is due to the zero-point motion since the Debye temperature for optical phonons is much larger than these temperatures ($\hbar\omega_{\text{phon}}/k_B \approx 570$~K). The cut-off remains unchanged when site-jitter is included, but all harmonics display a sharp intensity decrease. This is expected since, on the one hand, the breaking of periodicity opens small gaps in the band structure and leads to a quenching of intraband harmonics and, on the other, it leads to a loss of long-distance coherence between sites which affects interband harmonics.

To illustrate the latter, in Fig.~\ref{fig:gabor}a,b we plot the time-frequency maps of the finite-chain system without and with site jitter, that is, those corresponding to the gray and orange curves in Fig.~\ref{fig:hhg_finite}, respectively. In the system without site-jitter (Fig.~\ref{fig:gabor}a), we observe multiple trajectories per cycle, which produce the noisy high harmonic spectrum in Fig.~\ref{fig:gabor}d (gray curve). When optical phonon jitter is accounted for (Fig.~\ref{fig:gabor}b), we observe the typical short and long trajectory features~\cite{Vampa2015}. Such quenching of long-distance coherences and removal of long trajectories can be modeled by including a dephasing term which depends on the distance between the sites~\cite{Brown2024},
\begin{equation}
\begin{split}
    \mathcal{L}[\rho_{\alpha n,\beta n'}] =&  \gamma(|r_{\alpha,n} - r_{\beta,n'}|,x_{cut},\beta) \times \\
    &\left(\rho_{\alpha n, \beta n'} - \rho^{(0)}_{\alpha n, \beta n'} \right),
\end{split}
\end{equation}
where $\rho_0$ is the initial density matrix and $\gamma(x,x_{cut},\beta)$ is a function of distance,
\begin{equation}\label{eq:real_space_dephasing}
    \gamma(x;x_{cut},\beta) = 
\begin{cases}
    \beta (|x-x_{cut}|)^\alpha\,, \quad |x| \geq |x_{cut}| \\
    0 \,, \quad \text{otherwise}.
\end{cases}
\end{equation}
Fig.~\ref{fig:gabor}c shows the time-frequency plot for the same static 1D chain as in Fig.~\ref{fig:gabor}a, but including dephasing $\gamma(x;x_{cut},\beta)$. We used a linear site-distance dependence ($\alpha=1$) since we observed that it compares slightly better to the decoherence induced by optical phonons in our particular case. The time-frequency plot is similar to that with optical phonon jitter, and so is the high-harmonic spectrum (cf. orange and blue curves in Fig.~\ref{fig:gabor}d).

\begin{figure}
    \centering
    \includegraphics[width=\linewidth]{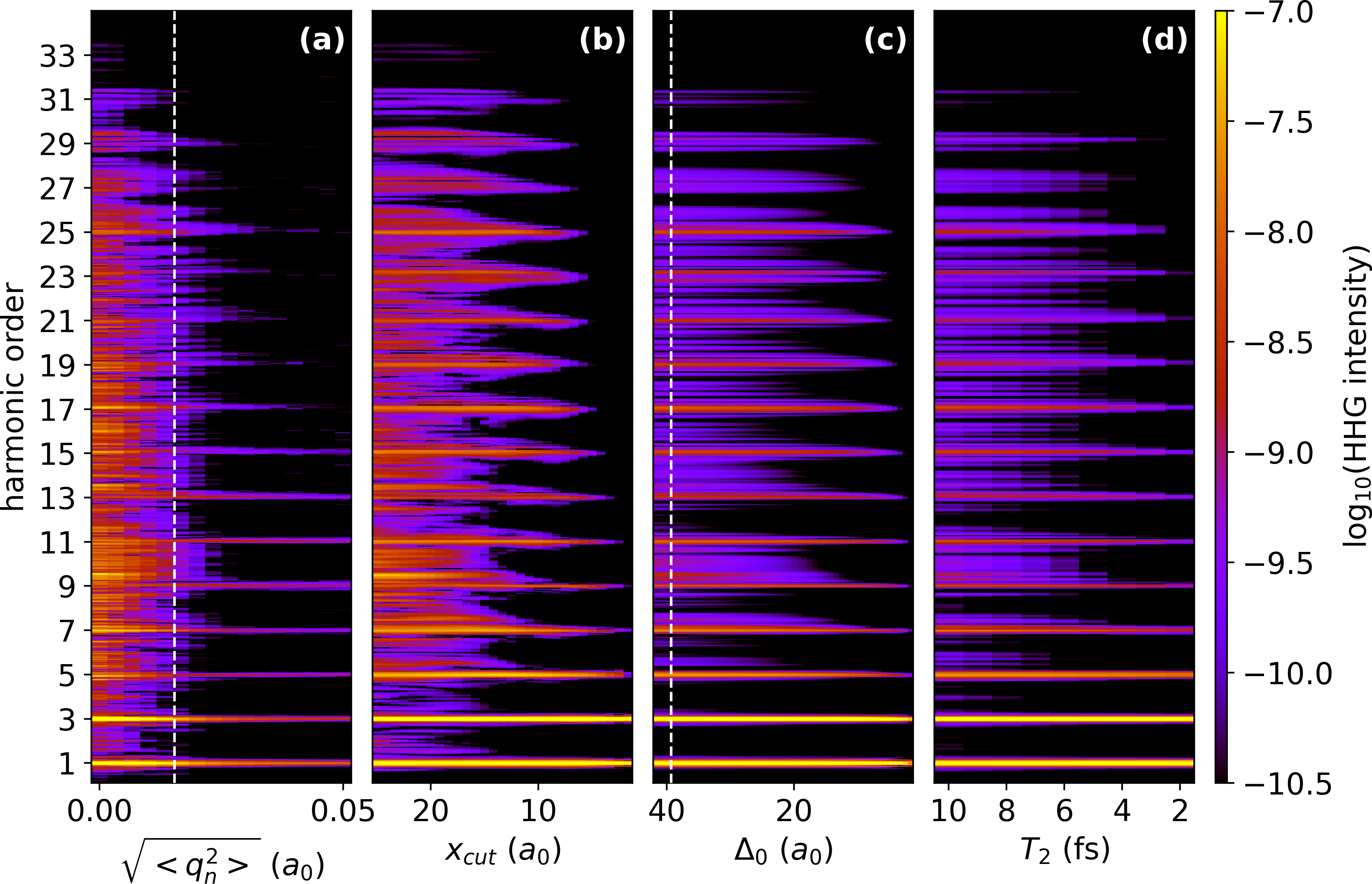}
    \caption{HHG spectra of a chain of $N=1199$ atoms as a function of different dephasing parameters: (a) root-mean-square of atomic displacement for a system with optical-phonon per-site jitter, (b) $\gamma(x; x_{cut},\beta)$ for fixed $\beta=2.4\times 10^{-3}$~a.u. and varying $x_{cut}$, (c) $\gamma(x; x_{cut},\beta)$ for fixed $x_{cut}=2a_0/3$ and varying $\Delta_0$, which corresponds to the distance for which the effective dephasing times is one quarter of the laser cycle, (d) dephasing time $T_2$. The white dashed line in panels (a,c) corresponds to the blue and orange HHG spectra in Fig.~\ref{fig:gabor}e, respectively.}
    \label{fig:scan}
\end{figure}

In order to explore more quantitatively the correspondence between the relative random displacement $q_n$ caused by optical phonons, and the ad-hoc distance-dependent dephasing $\gamma$ and dephasing time $T_2$ typically introduced in periodic calculations, in Fig.~\ref{fig:scan} we compare the HHG spectra as a function of these variables. In particular, for the case of the distance-dephasing $\gamma$, we change both the $x_{cut}$ parameter while keeping $\beta=2.4\times10^{-3}$~a.u. (Fig.~\ref{fig:scan}b), and keep $x_{cut}$ fixed while changing the distance $\Delta_0$ for which the effective dephasing time corresponds to one quarter of the laser cycle (Fig.~\ref{fig:scan}c). We find that, while neither $\gamma$ (Fig.~\ref{fig:scan}b,c) or $T_2$ (Fig.~\ref{fig:scan}d) are able to fully reproduce the effect of site-jitter in the HHG spectra (Fig.~\ref{fig:scan}a), they do reproduce the clean-up of the harmonic spectrum and the slight suppression of higher harmonics as decoherence is increased. They also reproduce the frequency regions where the spectrum is noisier when the dephasing rate is low (e.g., see region of harmonics H9-H11). They both fail to reproduce the drop in the yield of the lower-order harmonics. In fact, as $T_2$ is decreased, the yield of lower-order harmonics is increased due to dephasing-induced ionization~\cite{Boroumand2025,Molinero2025}, contrary to what is observed under increased optical phonon jitter.

We next consider local strain fluctuations generated by acoustic phonons. In contrast to optical modes, these excitations create long-wavelength, spatially correlated bond distortions. Details of the implementation can be found in Appendix ~\ref{sec:methods}. As before, we choose the parameters such that they are similar to those of MoS$_2$. In particular, we adjust the standard deviation of the bond stretch or compression parameter $\eta_n$ so that the detrended RMS atomic displacements, extracted from many disorder realizations, satisfy $\sqrt{\langle q^2 \rangle}=0.15$~\AA, a typical value for thermal acoustic phonons~\cite{Mannebach2015}, and we explore correlation lengths in the range $\ell_{corr} = 1-10$~nm, consistent with the characteristic thermal acoustic phonon wavelengths in MoS$_2$ at room temperature.

\begin{figure}
    \centering
    \includegraphics[width=\linewidth]{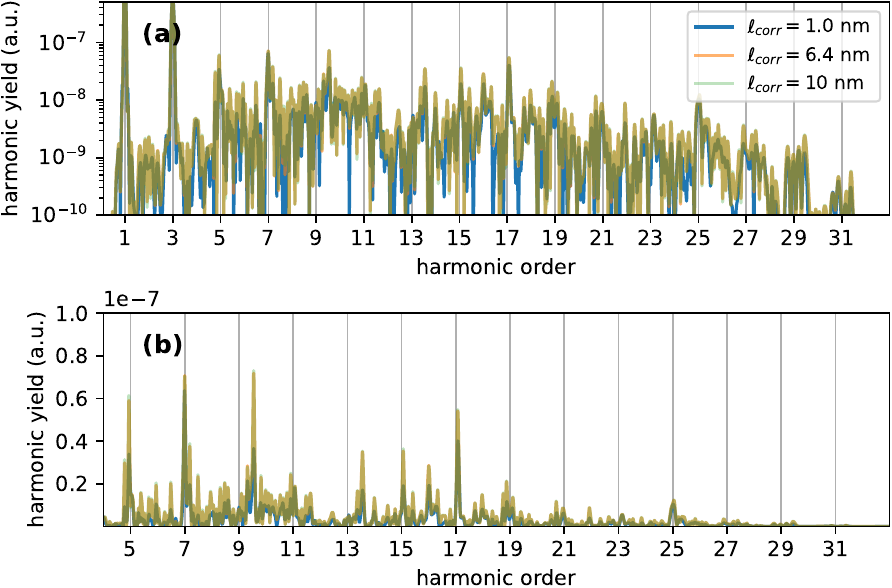}
    \caption{Effect of acoustic-phonon-like correlated jitter. HHG spectrum of a chain of $N=1199$ atoms with a correlated bond jitter with mean square atomic fluctuation $\langle q^2\rangle=0.15$~\AA\, (three times larger than that of optical phonons) and different correlation lengths $\ell_{corr}$. For all $\ell_{corr}$, spectrum is essentially the same as the frozen chain in Fig.~\ref{fig:hhg_periodic}, with negligible noise reduction. (a) Logarithmic scale. (b) Linear scale.}
    \label{fig:strain}
\end{figure}

Fig.~\ref{fig:strain} shows the resulting HHG spectrum. Although acoustic phonons generate larger RMS displacements than optical phonons, their nm-scale correlations do not have a strong impact in the HHG spectrum. The electron excursion length in the solid is only a few \AA\, - typically smaller than $\ell_{corr}$ - so electrons sample a nearly uniform local strain. As a result, the fine structure (or “noise”) of the HHG spectrum remains essentially unchanged relative to the unstrained system (Fig.~\ref{fig:hhg_periodic}), and is largely insensitive to realistic variations in $\ell_{corr}$, aside from minor overall intensity variations. The same reasoning applies to static local strain fields, such as those produced by impurities or Moiré corrugation in 2D materials, which typically have amplitudes of 0.1-0.3~\AA\,and nm-scale correlations~\cite{Nie2012,Schwarz2017}.

\begin{figure}
    \centering
    \includegraphics[width=\linewidth]{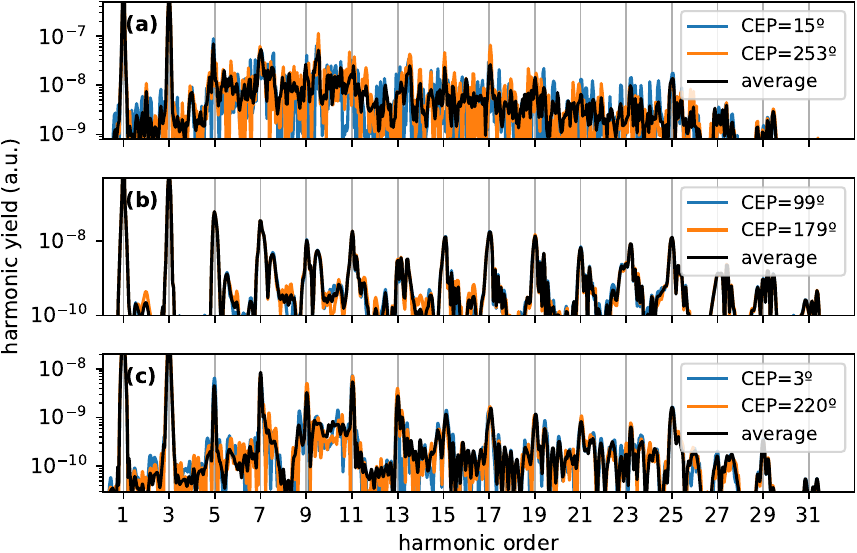}
    \caption{Long-pulse CEP-dependence of HHG spectra in a chain of $N=1199$ atoms. All panels show the spectra for two random CEP values (blue and orange lines) and the incoherent average of 50 random CEPs (black lines). (a) Periodic chain with no jitter and no dephasing. (b) Periodic chain with no jitter and site-distance-dependent dephasing $\gamma(x,x_{cut}=17~a_0)$. (c) Chain with optical-phonon-like per-site jitter. In c, the CEP average is taken by computing the coherent numerical average of the non-linear currents from 20 random nuclear configurations for one CEP value and then performing the incoherent numerical averaging of those average currents for 50 different CEP values. The pulse duration in all cases is 80~fs, corresponding to 22 cycles.}
    \label{fig:cep}
\end{figure}

Finally, we point out that the calculations so far were performed for a carrier-envelope-phase (CEP)-stable pulse. However, this is often not the case in HHG experiments. In cases were it is not, incoherent averaging over many CEP already introduces a slightly more defined harmonic spectrum. Fig.~\ref{fig:cep}a shows the HHG spectrum for two different CEP and for a random incoherent average of 50 CEP in a pristine, periodic chain. The pulse has a Gaussian envelope of 80~fs of FWHM, supporting 22 laser cycles, and hence one would expect a negligible effect of the CEP. 
However, the spectra differ strongly for different CEP.
Even if we assume that interband recombination in solids, as in the atomic three-step model, occurs within a fraction of an optical cycle, the interband coherence created at each excitation event can persist for many cycles in a pristine crystal. This long-lived coherence acts as a memory reservoir that allows short trajectories launched in different cycles to be coupled and interfere, giving rise to non-adiabatic CEP-dependent emission even for long pulses. This effect can alternatively be viewed as originating from the population of multiple Floquet states~\cite{Lange2024}. When either distance-dependent dephasing or optical-phonon-induced site jitter is included (Fig.~\ref{fig:cep}b,c), this long-range interband coherence is quenched, decoupling the sub-cycle trajectories across the different cycles. As a result, the HHG process becomes effectively adiabatic and CEP insensitive for long multi-cycle pulses. Hence, CEP-stable HHG experiments in solids may provide a way to retrieve RMS deviations from pristine atomic positions and, more generally, the coherence lengths relevant in solid-HHG.

\section{Conclusions\protect\label{sec:conclusions}}
In conclusion, we have shown that atomic-site-jitter consistent with optical phonon amplitudes leads to a loss of long-distance coherence and thus to the kind of clean HHG spectra typically observed in the experiments. Such effect can be qualitatively reproduced by using a site-distance-dependent dephasing $\gamma$, recently proposed in the context of solid-state HHG~\cite{Brown2024}. In contrast to optical phonons, where the atomic displacements can be considered random per-site, acoustic phonons and, in general, local bond strain, produce site jitter that is correlated over several nanometers. This is comparable or even larger than the electron quiver radius in the solid and hence the effect of local, correlated strain fluctuations, like those induced by thermal effects, will generally not impact the HHG spectrum considerably. Finally, we have discussed the effect of CEP-instability in the harmonic spectrum, finding that, for realistically long pulses, CEP effects are present if distant sites maintain coherence. As soon as this long-distant coherence is lost, for example by the effect of optical phonons, CEP-dependent effects reduce dramatically. In this way, information on electron zero-point motion and coherence may be retrieved with a CEP-controlled HHG experiment.

\section*{Acknowledgments}

The authors thank Thomas Hansen for useful discussions. This work was supported through the Talento Comunidad de Madrid Fellowship 2022-T1/IND24102 and the Spanish Ministry of Science, Innovation and Universities through grant references PID2023-146676NA-I00, PID2021-122769NB-I00, RYC2022-035373-I and PID2024-161665NB-I00, funded by MICIU/AEI/10.13039/501100011033, “ERDF A way of making Europe” and “ESF+”. D. Purschke acknowledges the Mitacs Globalink Research Award and the NSERC Postdoctoral Fellowship Program.

\appendix

\section{Methods}\label{sec:methods}
We start by considering the full Hamiltonian for electrons with position operators $\hat{r}$ and nuclei with position operators $\hat{R}$,
\begin{equation}
    \hat{H}(t) = \hat{T}_n + \hat{H}_{el}(t;\hat{R})
\end{equation}
where the nuclear kinetic operator,
\begin{equation}
    \hat{T}_n = \sum_{j}\frac{\hat{P}_{j}^2}{2M_{j}}
\end{equation}
and the electronic Hamiltonian,
\begin{equation}
    \hat{H}_{el}(t;\hat{R}) = \hat{T}_e+\hat{V}_{nn}(\hat{R})+\hat{V}_{ee}+\hat{V}_{en}(\hat{R})+\hat{H}_{int}(t;\hat{r}),
\end{equation}
is composed of the electronic kinetic term, the nuclear and electronic potentials, the electron-nuclei interaction and the interaction Hamiltonian with the external laser field. The equation of motion for the full density matrix follows the von Neumann equation (atomics units are used unless otherwise stated)
\begin{equation}
    i\partial_t \hat{\rho}(t) = [\hat{H}(t),\hat{\rho}(t)].
\end{equation}
In the interaction picture relative to the nuclear kinetic term,
\begin{equation}
    \hat{H}_I (t) = e^{i \hat{T}_n t}\,\hat{H}_{el}(t;\hat{R})\,e^{-i\hat{T}_n t}
\end{equation}
with $    \hat{\rho}_I(t) = e^{i\hat{T}_n t}\,\hat{\rho}(t)\,e^{-i\hat{T}_n t}
$. 

The laser-driven electronic dynamics relevant for HHG occurs on a much shorter timescale than any appreciable nuclear motion. Therefore, we can assume that the nuclear kinetic energy can be neglected over the pulse duration $\tau$, i.e., that the nuclei move only a small distance compared to the interatomic spacing, $\frac{P}{M}\,\tau \ll a_0$, with $a_0$ the inter-atomic distance. Then, $e^{i \hat{T}_n t} \approx 1$ over the pulse, so that $\hat{H}_I \approx \hat{H}_{el}(t;\hat{R})$, i.e., the electronic Hamiltonian depends only on $\hat{R}$ and not on the nuclear momenta $\hat{P}$. Hence, the nuclear coordinates become constants of motion,
\begin{equation}
    [\hat{R},\hat{H}(t)] = 0,
\end{equation}
and $R$ can be treated as a static parameter. This is the ``frozen-nuclear-wavepacket'' approximation, which is particularly applicable in the context of HHG due to the intrinsic fast timescales~\cite{Hansen2017}. The quantum probability density of the nuclear wavepacket is
\begin{equation}
    P(R,t) = \sum_{\alpha}\rho_{\alpha,\alpha}(t;R,R),
\end{equation}
where $\alpha$ labels the electronic basis. Since the nuclear coordinates are frozen when $\hat{T}_n=0$, we have
\begin{equation}
    P(R,t) \xrightarrow{\hat{T}_n = 0} \sum_{\alpha}\rho_{\alpha,\alpha}(t=0;R,R) \equiv P(R).
\end{equation}
We can then define the time-dependent electronic density matrix for a particular nuclear configuration as
\begin{equation}\label{eq:frozen_nuclei}
    \hat{\rho}_{el} (t;R) = \frac{\hat{\rho}(t;R,R)}{P(R)},
\end{equation}
which follows the equation of motion
\begin{equation}
    i \partial_t \hat{\rho}_{el}(t;R) = [\hat{H}_{el}(t;R), \hat{\rho}_{el}(t;R)].
\end{equation}
In general the full density matrix is $\hat{\rho} (t;R,R')$ and may contain coherences between different nuclear coordinates $R\neq R'$. To compute the HHG spectrum, however, it is sufficient to focus only in the cases when $R=R'$. This is because we are computing an electronic observable, i.e., the expectation value of the current,
\begin{equation}
    \langle J(t) \rangle 
    = \mathrm{Tr}_{R,el}[\hat{J}\,\hat{\rho}(t)],
\end{equation}
which upon tracing out the nuclear degrees of freedom becomes
\begin{equation}
    \langle J(t) \rangle 
    = \int dR\,\mathrm{Tr}_{el}[\hat{J}\,\hat{\rho}(t;R,R)].
\end{equation}
Using the ``frozen-nuclear-wavepacket" approximation in Eq.\ref{eq:frozen_nuclei}, we have
\begin{equation}
    \langle J(t) \rangle 
    = \int dR\, P(R)\, J_R(t),
\end{equation}
where
\begin{equation}
    J_R(t) = \mathrm{Tr}_{el}[\hat{J}\,\hat{\rho}_{el}(t;R)].
\end{equation}
This ensemble average can be obtained numerically by averaging over several random realizations 
$R^{(k)}$ drawn from $P(R)$,
\begin{equation}
    \langle J(t) \rangle 
    \approx \frac{1}{N_{\mathrm{seeds}}}
    \sum_{k=1}^{N_{\mathrm{seeds}}} J_{R^{(k)}}(t).
\end{equation}
If the system contains a sufficiently large number of atoms, only a few random seeds are required for convergence due to self-averaging effects.

We construct the field-free electronic Hamiltonian $\hat{H}_{el}(t=0;R)$ with a tight binding model of a 1D semiconductor, modeled by a tight binding diatomic chain with alternating A and B atoms and random distances between them,
\begin{equation}
\begin{split}
    \hat{H}_{el}(t=0) &= \sum_n \Big[ \varepsilon_A\,a_n^\dagger a_n + \varepsilon_B\,b_n^\dagger b_n \\
    & -t(d_{1,n}) a_n^\dagger b_n - t(d_{2,n}) a_{n+1}^+ b_n + \text{h.c.}\Big],
\end{split}
\end{equation}
where $a_n^+ (a_n)$ and $b_n^+ (b_n)$ are creation/annihilation operators on sites A and B, respectively, of cell $n$ and $\varepsilon_{A/B}$ are the atomic on-site energies. The intracell and intercell nearest neighbour distances for the pristine (no jitter) case are, respectively,
\begin{equation}
\begin{split}
    &d_{1,n}^{(0)} = x_B-x_A \\
    &d_{2,n}^{(0)} = a_0 - (x_B-x_A),
\end{split}
\end{equation}
while in the ``jittered" case,
\begin{equation}
\begin{split}
    &d_{1,n} = r_{B,n} - r_{A,n} \\
    &d_{2,n} = r_{A,n+1} - r_{B,n},
\end{split}
\end{equation}
where $r_{A,n}$ and $r_{B,n}$ are defined below for each model. The distance-dependent hopping is modeled as~\cite{Pereira2009},
\begin{equation}
    t(d) = t_0\,e^{-\gamma(d-d^{(0)})/d^{(0)}},
\end{equation}
where $t_0$ is the hopping amplitude for the pristine crystal, $\gamma$ is the distance decay rate and $d$ is modified by the jitter.

\subsection{Optical phonon jitter}
To model jitter produced by optical phonons, we position atoms A and B on cell $n$ at
\begin{equation}
\begin{split}
    &r_{A,n} = n a_0 + x_A+s_A \,q_n \\
    &r_{B,n} = n a_0 + x_B-s_B \,q_n, \\
\end{split}
\end{equation}
where $a_0$ is the lattice constant, $s_{A/B} = \frac{m_{A/B}}{m_A+m_B}$ are mass factors that scale the atomic displacement and $q_n$ is the random displacement in cell $n$, which we model as a Gaussian with mean-square~\cite{RichardPFeynman1998},
\begin{equation}
    \langle q_n^2 \rangle = \frac{\hbar}{2\mu \omega}\coth(\frac{\hbar \omega}{2k_BT}).
\end{equation}
In terms of $R=q_n$ we approximate $P(R) \equiv P(\{q_n\})\approx \prod_n P(q_n)$, which corresponds to neglecting spatial correlations of optical zero-point motion. We take $\omega$ as the optical phonon frequency of the material at wavevector $\Gamma$, $\mu$ is the effective mass, $k_B$ is the Boltzmann constant and $T$ is the temperature.

\subsection{Acoustic phonon / local strain jitter}
To model jitter produced by acoustic phonons or local strain, the positions of atoms A and B on cell $n$ are
\begin{equation}
\begin{split}
    &r_{A,n} = n a_0 + x_A+q_n \\
    &r_{B,n} = n a_0 + x_B+q_n. \\
\end{split}
\end{equation}
Here, we sample the probability $P(R)$ by assigning to each bond a small random stretch or compression, $\eta_n$, drawn from a zero-mean Gaussian distribution with adjustable standard deviation. We then impose acoustic-like coherence by smoothing this noise through convolution with a Gaussian function of width $\ell_{\mathrm{corr}}$,
\begin{equation}
\delta_n = \sum_m g_{n-m}\eta_m, \qquad g(x) \propto e^{-x^2/(2\ell_{\mathrm{corr}}^2)}.
\end{equation}
The resulting $\delta_n$ field varies smoothly over distances set by $\ell_{\mathrm{corr}}$, mimicking the strain profile of a long-wavelength phonon. After smoothing, we subtract the spatial average of $\delta_n$ to enforce zero-mean random strain. Atomic displacements are then obtained by cumulatively summing the bond distortions,
\begin{equation}
q_n = \sum_{i \le n} \delta_i,
\end{equation}
such that $P(R)\equiv P(\{q_n\})$ in this case is a zero-mean correlated Gaussian distribution with its amplitude and spatial correlations controlled by the standard deviation of $\eta_n$ and $\ell_{corr}$.

\bibliographystyle{apsrev4-1}
\bibliography{biblio}

\end{document}